\documentclass[fleqn,usenatbib]{mnras}

\usepackage{newtxtext,newtxmath}

\usepackage[T1]{fontenc}
\usepackage{ae,aecompl}

\usepackage{graphicx}	
\usepackage{amsmath}	
\usepackage{amssymb}	
\usepackage{xcolor}
\usepackage[normalem]{ulem}

\hypersetup{draft}

\newcommand{\Teff}{\mbox{$T_{\rm eff}$}}

\newcommand{\kepler}{{\em Kepler\/}}
\newcommand{\dsct}{\mbox{$\delta$~Sct}}
\newcommand{\gdor}{\mbox{$\gamma$~Dor}}
\newcommand{\cd}{\mbox{d$^{-1}$}}

\DeclareMathOperator{\sinc}{sinc}

\title[Period--luminosity relation for $\delta$ Scuti stars]{The period--luminosity relation for $\delta$~Scuti stars using {\em Gaia} DR2 parallaxes}

\author[Ziaali et al.]{%
Elham Ziaali$^{1,2}$\thanks{E-mail: e-ziaali@riaam.ac.ir}, 
Timothy R. Bedding$^{2,3}$\thanks{E-mail: tim.bedding@sydney.edu.au},
Simon J. Murphy$^{2,3}$, \newauthor
Timothy Van Reeth$^{2,3}$ and
Daniel R. Hey$^{2,3}$
\\
$^1$Research Institute for Astronomy and Astrophysics of Maragha (RIAAM), PO Box 55134-441, Maragha, Iran\\
$^2$Sydney Institute for Astronomy, School of Physics, University of Sydney 2006, Australia \\
$^3$Stellar Astrophysics Centre, Department of Physics and Astronomy, Aarhus University, 8000 Aarhus C, Denmark
}

\date{Accepted 16 April 2019.}

\pubyear{2019}

\begin{document}
\label{firstpage}
\pagerange{\pageref{firstpage}--\pageref{lastpage}}
\maketitle

\begin{abstract}
We have examined the period--luminosity (P--L) relation for $\delta$ Scuti stars using {\em Gaia} DR2 parallaxes. We included 228 stars from the catalogue of Rodriguez et al.\ (2000), as well as 1124 stars observed in the four-year \kepler\ mission. For each star we considered the dominant pulsation period, and used DR2 parallaxes and extinction corrections to determine absolute $V$ magnitudes. Many stars fall along a sequence in the P--L relation coinciding with fundamental-mode pulsation, while others pulsate in shorter-period overtones. The latter stars tend to have higher effective temperatures, consistent with theoretical calculations.  Surprisingly, we find an excess of stars lying on a ridge with periods half that of the fundamental.  We suggest this may be due to a 2:1 resonance between the third or fourth overtone and the fundamental mode.
\end{abstract}




\section{Introduction}

Period--luminosity (P--L) relations of pulsating stars have a long and distinguished history \citep{Leavitt+Pickering1912}. They arise when a class of pulsating stars occupies a relatively narrow range of effective temperatures, in which case luminosity correlates quite strongly with stellar density, and hence with the pulsation periods of pressure modes \citep[e.g.,][]{Eddington1926,Carroll+Ostlie2006}. In this paper, we investigate the P--L relation for the $\delta$~Scuti class of pulsating stars using parallaxes from {\em Gaia} DR2 \citep{Gaia2018}.

The \dsct\ stars are intermediate-mass stars with spectral types A2V to F2V that pulsate in low-order pressure modes. They are located within an interesting region in the Hertzprung-Russell (HR) diagram, between low-mass stars having thick convective envelopes ($\leq 1M_{\sun}$) and high-mass stars with large convective cores and radiative envelopes ($\geq 2M_{\sun}$). 
They lie in the lower part of the classical instability strip, within or just above the main sequence in HR diagram, with effective temperatures between approximately 6400\,K and 8600\,K and pulsation frequencies above 5\,d$^{-1}$, often with multiple periodicities \citep[e.g.,][]{Breger2000,Michel2017,Balona2018,Bowman+Kurtz2018,Pamyatnykh2000,Qian2018}.

In \dsct\ stars, pulsations are excited by the well-known $\kappa$ mechanism, which operates in zones of partial ionization of hydrogen and helium. In cooler \dsct\ stars with substantial outer convection zones, the selection mechanism of modes with observable amplitudes could be affected by induced fluctuations of the turbulent convection. 
A theoretical blue edge of the instability strip for radial and non-radial modes was determined by \citet{Pamyatnykh2000}. At the red edge, pulsation is damped by convection such that a non-adiabatic treatment of the interaction between convection and pulsation is required, in so-called time dependent convection (TDC) models. \citet{Houdek2000} and \citet{Xiong2001} used TDC to study the red edge of radial modes, while a red edge for non-radial modes was produced three years later by \citet{Dupret2004}. TDC has an adjustable mixing length parameter, $\alpha_{\rm MLT}$, whose value influences the position of both the blue and red edges. \citet{Dupret2005} calculated instability strips of radial and non-radial modes for different values of $\alpha_{\rm MLT}$.

Like RR~Lyraes and Cepheids, which also lie in the instability strip, the \dsct\ stars are known to follow a P--L relation, at least for high-amplitude pulsators and admittedly with considerable scatter \citep[e.g.,][]{Breger+Bregman1975,King1991,North1997, McNamara1997,Poretti2008,Garg++2010,Poleski++2010,McNamara2011,Cohen+Sarajedini2012}.
In particular, \citet{McNamara2011} studied the P--L relation of high-amplitude \dsct\ stars and used it to find the distance moduli of three galaxies and two globular clusters. 
In this paper we revisit the P--L relation for \dsct\ stars using {\em Gaia} DR2 parallaxes, considering only the highest-amplitude mode in each star.  It is important to keep in mind that \dsct\ stars can pulsate in fundamental modes ($n=1$) and also in overtone modes ($n=2, 3, 4, \ldots$), and that these modes can be either radial ($l=0$) or nonradial ($l=1, 2, \ldots$).  In general, if the strongest oscillation mode in a star falls on the P--L relation, it is likely to be the radial fundamental mode ($n=1$, $l=0$) or possibly a low-order dipole mode ($l=1$), while other modes are expected to have shorter periods.

In Section~2 we describe our sample's construction. In Section~3 the P--L diagrams are plotted, and possible explanations for the P--L relation are discussed in Section~4. 

\section{Construction of samples}
\label{sec:samples}

\subsection{Rodriguez et al. (2000) catalogue}

\citet{Rodriguez2000} catalogued the dominant pulsation period for 636 \dsct\ stars. Based on the list of rejected \dsct\ stars in Table~5 of \citet{Liakos+Niarchos2017}, three stars from the catalogue (BQ\,Phe, DE\,Oct, V345\,Gem) were identified as binaries with no pulsating components and removed from our sample. We also removed AK Men as a binary after light curve inspection. 
Fourteen stars from crowded parts of sky were removed because their coordinates were not precise enough to reliably query against dust maps \citep[e.g.][]{Green2015}. A further six stars (HD\,302013 = V753\,Cen, HD\,358431 = YZ\,Cap, TV\,Lyn, BP\,Peg, DH\,Peg, and UY\,Cam) were excluded for being RR Lyrae stars \citep[e.g.,][] {Sneden2018,McNamara2011}. We also dropped the B-type $\beta$\,Cep variable, V1228\,Cen \citep{Pigulski2008}. Two further stars without measured parallaxes were also discarded (HD\,23567 = V534\,Tau and 2MASS\,J13554648-291123).

\label{sec:additional}

We also included four bright \dsct\ stars that were recently discovered, and are therefore not in the \citet{Rodriguez2000} catalogue (shown by cyan triangles in Figs.~\ref{fig:PL} and~\ref{fig:PL_2}): 
\begin{itemize}
\item 
Altair ($\alpha$~Aql) was found to be the brightest \dsct\ star in the sky ($V=0.76$) by \citet{Buzasi2005} using WIRE photometry, with a dominant frequency of 15.77\,\cd. It is included in Fig.~1 of \citet{McNamara2011}.

\item $\beta$~Pic has a dominant frequency of 47.4\,\cd, based on photometry from Antarctica \citep{Mekarnia2017}.

\item 95~Vir has a dominant frequency of 9.537\,\cd, based on \kepler\ K2 photometry \citep{Paunzen2017}.

\item $\eta$~Ind (HR 7920) has a dominant frequency of 26.5\,\cd, although this could be affected by daily aliases \citep{Koen2017}.

\end{itemize}

We calculated the absolute magnitudes for all these stars using
\begin{equation} \label{abs_mag_rod}
M_{V} = V + 5\log\pi + 5 - A_V,
\end{equation}
where $M_{V}$ and $V$ denote the absolute and apparent magnitudes in the $V$ band, respectively, $\pi$~is the parallax in arcsec, and $A_V$~is the extinction due to interstellar dust. We used parallaxes from {\em Gaia} DR2 except for a few cases where the {\em Hipparcos} parallaxes were more precise.  We calculated $A_V$ using the dust map by \citet{Green2015}.

\subsection{\kepler\ \dsct\ stars}
\label{sec:kepler-sample}

Many \dsct\ stars were observed by \kepler\ during its four-year nominal mission \citep[e.g.,][]{Balona+Dziembowski2011,Balona2015,Bradley2015,Moya2017,Balona2018,Barcelo-forteza2018,Bowman+Kurtz2018}. We have used the sample of about 2000 \dsct\ stars identified by \citet{Murphy2019} from \kepler\ long-cadence data.  To account for the averaging of pulsations during the 29.4-minute integrations ($f_{\rm Nyq} = 24.5\,\cd$), we divided the observed amplitudes in the Fourier spectrum by the function $\sinc \left(\frac{\pi}{2} \frac{f}{f_{\rm Nyq}}\right)$ \citep[e.g.,][]{Huber2010}. 
We measured the dominant period in each star from the highest peak in the Fourier spectrum in the range of 0.0228 to 0.2 days (i.e., with frequencies from 5 to 43.9\,\cd).  

We calculated the absolute magnitudes for the stars in the \kepler\ sample using
\begin{equation} \label{abs_mag_kepler}
M_{V} = V - 5\log d + 5 - A_V,
\end{equation}
where $M_{V}$ and $V$ denote the absolute and apparent magnitudes in the $V$ band, respectively, $d$~is the distance in pc, and $A_V$~is the extinction due to interstellar dust. Apparent magnitudes were obtained from \citet{Everett2012}, accessed via MAST\footnote{Mikulski Archives for Space Telescopes \texttt{https://archive.stsci.edu/}} (V\_UBV) and assigned an uncertainty of 0.05 mag. To obtain stellar distances from Gaia DR2 parallaxes, we used the normalised posterior distribution and adopted length scale model of \cite{Bailer-Jones2018Estimating}. This approach is advantageous, as it allows for a Bayesian approach to distance estimation. This produces a distribution of distances from which Monte Carlo samples can be drawn.
Extinctions and their uncertainties were obtained with the  \textsc{dustmaps} Python package \citep{Green-dustmaps2018}, which provides access to the Bayestar 17 reddening map of \cite{Green2018}. To convert to the appropriate photometric system (Johnson V), we used the extinction coefficient in Table A1 of \cite{Sanders2018Isochrone}. A small recommended grey offset of 0.063 was introduced into the extinctions. We determined the magnitude and its associated uncertainty for each star with a Monte-Carlo process, taking the mean value of 200 000 samples for the magnitude, and the 1-$\sigma$ standard deviation for the uncertainty.

\section{Period--Luminosity Diagram}
\label{sec:method}

To study the P--L relation with high accuracy, we preferentially chose stars with good parallaxes and reasonably small extinction corrections.  For the \citet{Rodriguez2000} catalogue, we used 228 stars with $V<14$ mag, fractional parallax uncertainties less than 5 percent (translating to uncertainties less than 0.11\,mag in $M_V$) and $A_V < 0.2$.  For the \kepler\ sample, we used 1124 stars with fractional parallax uncertainties less than 5 percent and $A_V<0.5$. 

It is known that some \dsct\ stars show amplitude variability \citep[see][and references therein]{Bowman2016}, even to the extent that the dominant mode (that is, the mode having the highest amplitude) might vary with time.  This would introduce some horizontal scatter in the P--L diagram.
Meanwhile, the light variations in high-amplitude \dsct\ stars (HADS) can reach a few tenths of a magnitude, which could
lead to some additional scatter in the vertical direction.

\begin{figure}
\includegraphics[width=1.0\linewidth]{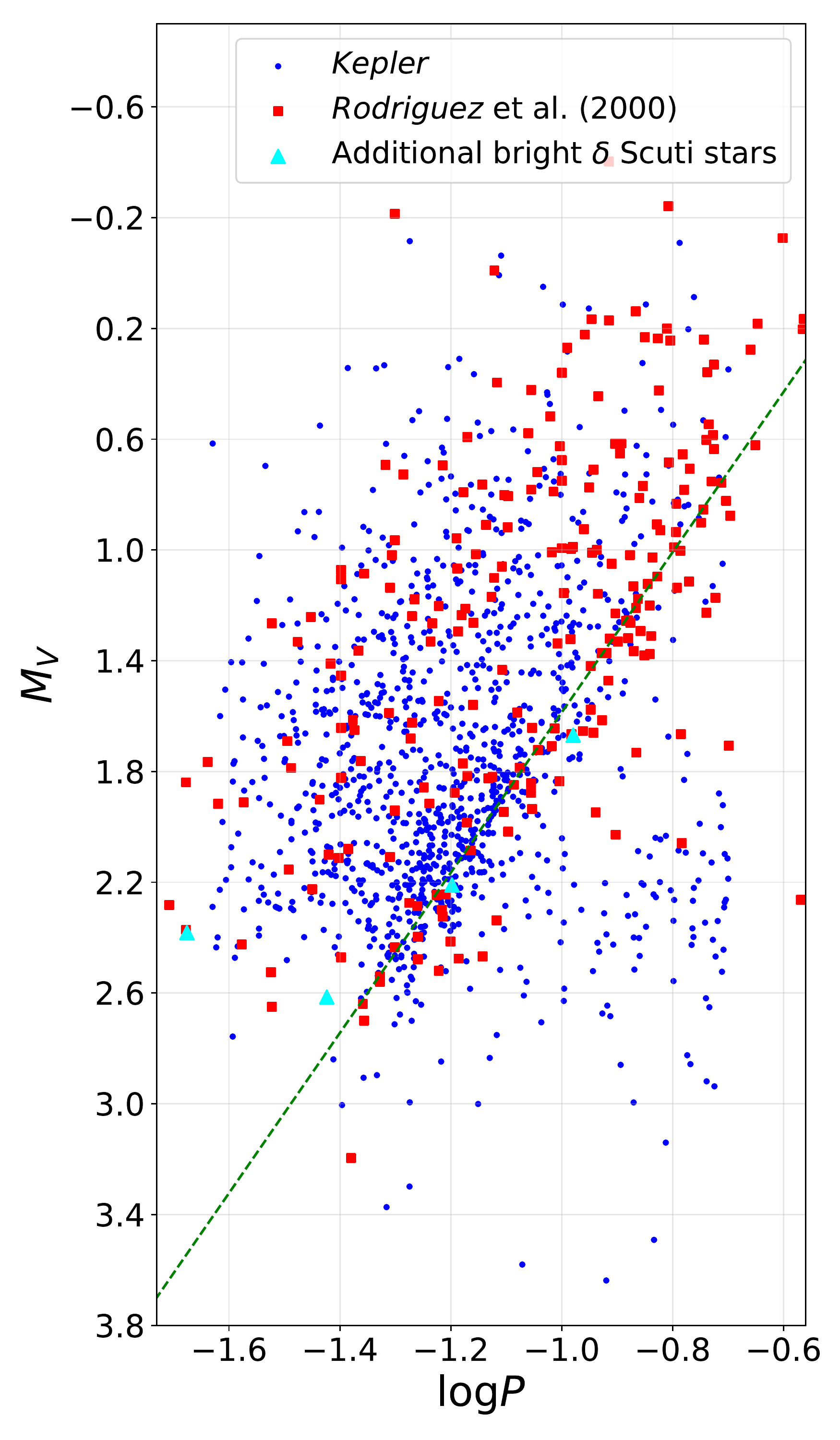}
\caption{Period--luminosity diagram of \dsct\ stars having fractional parallax uncertainties less than 5 percent. Blue dots are 1124 \dsct\ stars in the \kepler\ field and red squares are 228 stars from the \dsct\ catalogue of \citet{Rodriguez2000}. The cyan triangles indicate the four additional bright \dsct\ stars discussed in Sec.~\ref{sec:additional}, which are (from left to right) $\beta$~Pic, $\eta$~Ind, Altair and 95~Vir. The diagonal dashed green line shows the relation from \citet[][see Eq.~\ref{eq:McNamara}]{McNamara2011}. }
\label{fig:PL}
\end{figure}

\begin{figure}
\includegraphics[width=1.0\linewidth]{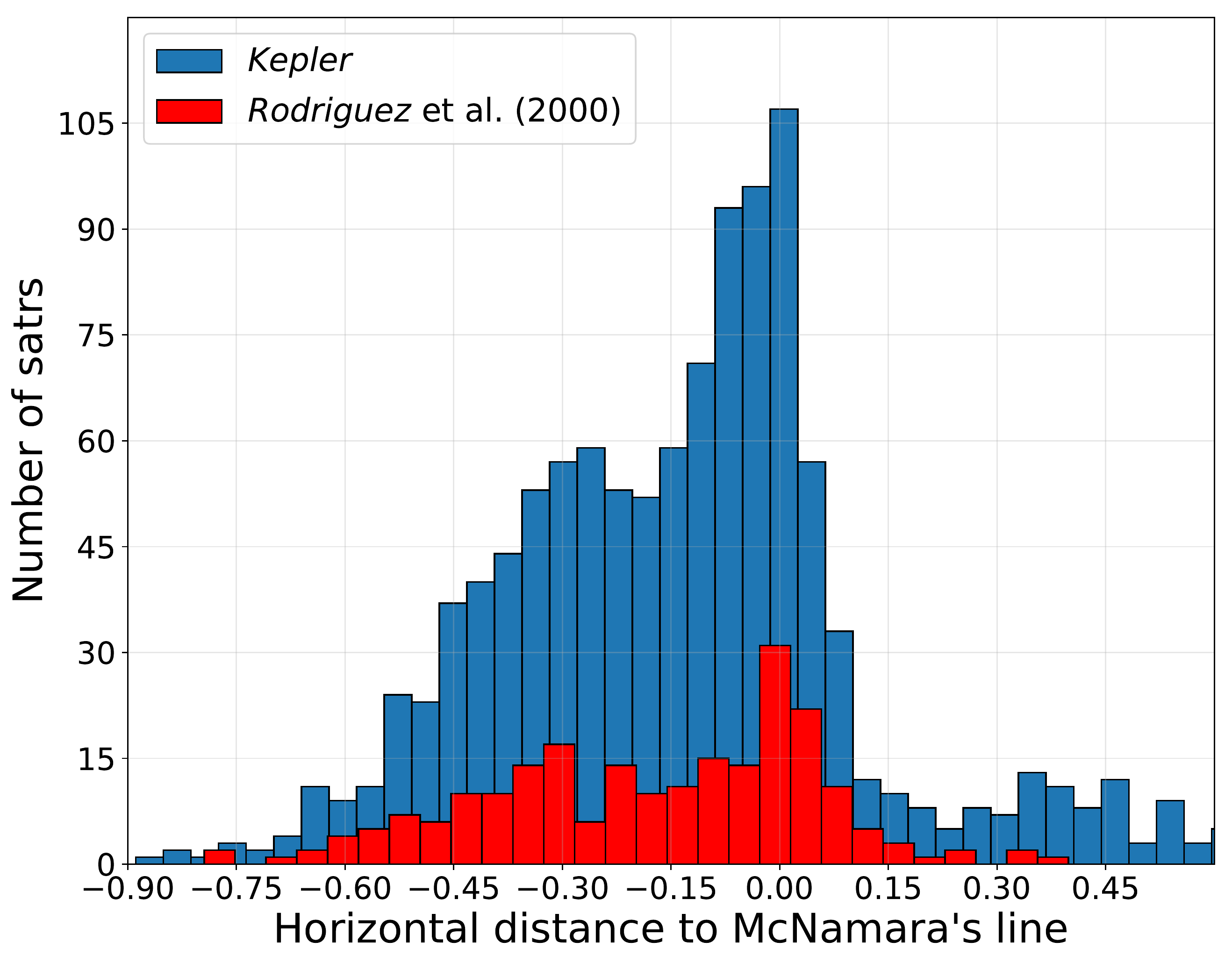}
\caption{The distance of data points to the McNamara line as a histogram. 
Red: 228 \dsct\ stars from the \citeauthor{Rodriguez2000} catalogue corrected for extinction by the \citet{Green2015} dust map.
Blue: 1124 \dsct\ stars from Kepler, corrected for extinction by the \citet{Green2018} dust map.}
\label{fig:histogram}
\end{figure}

Figure~\ref{fig:PL} shows our resulting P--L relation.
Red squares are stars from \citet{Rodriguez2000} catalogue, blue circles are \textit{Kepler} stars and cyan triangles are the additional stars mentioned in \ref{sec:additional}. The diagonal green line is the P--L relation derived by \citet{McNamara2011} for the high-amplitude \dsct\ stars, which are thought to pulsate in their radial fundamental mode.

We see that many stars fall in a ridge close to the dashed line.  Note that the group of \kepler\ stars in the lower right of Fig.~\ref{fig:PL}, whose periods appear to be too long for \dsct\ pulsations, are mostly \gdor\ stars with harmonics above 5\,d$^{-1}$ in the Fourier spectrum \citep[see][]{Murphy2019}.

Many stars in Fig.~\ref{fig:PL} lie to the left of the fundamental ridge, indicating that their dominant pulsation period corresponds to an overtone mode. Surprisingly, there appears to be an excess of stars in a second ridge that lies to the left of the main ridge.  
To make this clearer, we show in Fig.~\ref{fig:histogram} the histogram of horizontal distances from the fundamental-mode ridge.  That is, the red histogram shows the horizontal distances (in $\log P$) from the \citet{McNamara2011} line of stars in the \citet{Rodriguez2000} sample, while the blue histogram shows the same for the \kepler\ stars. The main peak (at zero distance in $\log P$) corresponds to the fundamental-mode ridge.  The second peak is clearly visible in both samples, and indicates that this ridge is displaced horizontally by 0.3 in $\log P$, corresponding to a period ratio of ($10^{-0.3}$\,=)\,0.50.  Thus, it appears that a significant number of stars in both samples have a dominant pulsation period that is half that of the fundamental mode.

\begin{figure}
\includegraphics[width=1.0\linewidth]{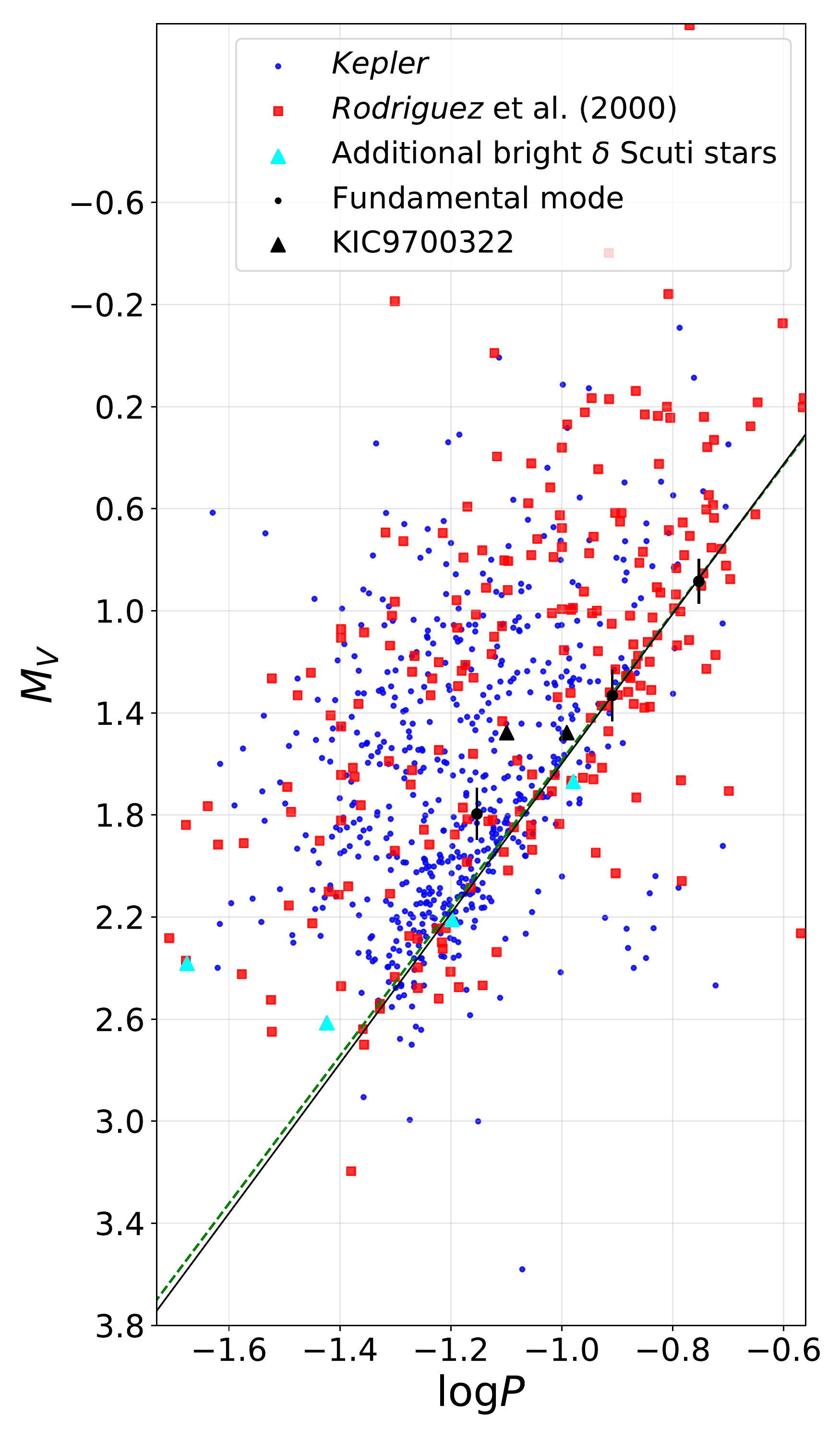}
\caption{Same as Fig.~\ref{fig:PL}, except that the \kepler\ sample (blue dots) is restricted to 601 stars having a pulsation semi-amplitude above 1\,mmag, and an offset of 0.125\,mag has been subtracted from $V$ magnitudes for the \kepler\ sample (see text).   The diagonal lines show the McNamara relation (dashed green line; see Eq.~\ref{eq:McNamara}) and our new fit (solid black line; see Eq.~\ref{eq:newfit}). Three \kepler\ stars for which the dominant mode has been identified as the radial fundamental are marked by black circles, with error bars.
From left to right, they are KIC~5950759, KIC~2304168 and KIC~9408694 (V2367~Cyg).  The black triangles show the two strongest modes in KIC~9700322, the left one corresponding to the first radial overtone (the dominant mode) and the right one being the fundamental radial mode \citep{Breger2011}}.
\label{fig:PL_2}
\end{figure}

Our two samples are somewhat different, given that the \citet{Rodriguez2000} catalogue comprises ground-based data, often from relatively short observations, whereas \kepler\ observed from space for four years.  To make the comparison more similar, we plotted the histogram after restricting the \kepler\ sample to stars with semi-amplitudes (measured in the Fourier transform) above 1\,mmag.  The result is shown in Fig.~\ref{fig:PL_2}.   
The histograms for the two samples do indeed appear to be more similar, although there are still some differences. The main point is the two ridges separated by a factor of two in period appear in both samples.

\begin{figure}
\includegraphics[width=1.0\linewidth]{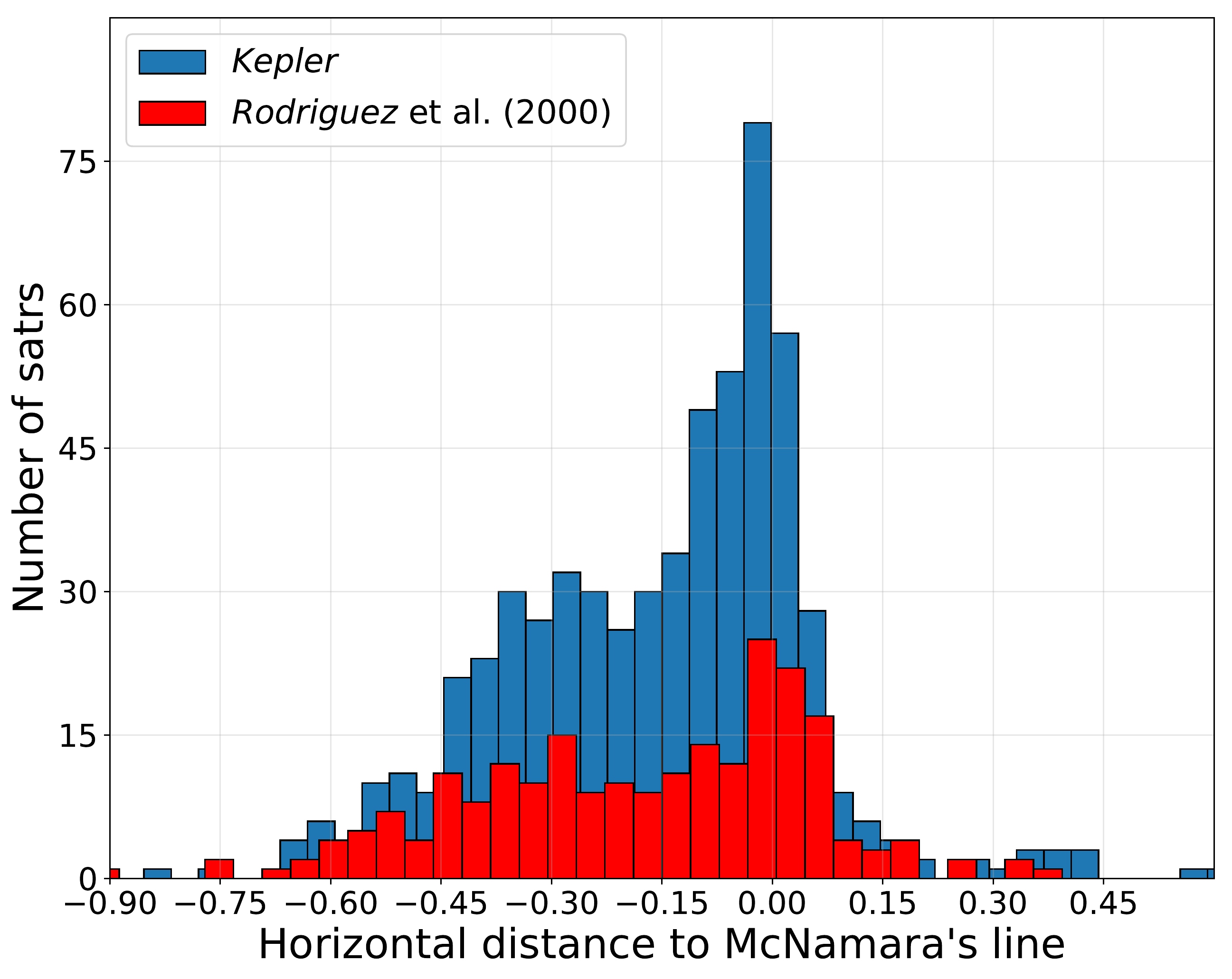}
\caption{Same as Fig.~\ref{fig:histogram}, except applied to the P--L diagram in Fig.~\ref{fig:PL_2}.}
\label{fig:histogram-highamp-offset}
\end{figure}

A fit to the P--L relation for a small number of \dsct\ stars was made by \citet{McNamara2011}, who found that
\begin{equation} \label{eq:McNamara}
M_V = (-2.89 \pm 0.13) \log P - (1.31 \pm 0.10).
\end{equation}
This is shown as the green dashed line in Figs.~\ref{fig:PL} and~\ref{fig:PL_2}. In order to determine a revised equation for this P--L relation, we fitted to all stars in both samples that are located on the fundamental ridge.
The fitted line is shown by the black solid line in Fig.~\ref{fig:PL_2} and has the following equation:
\begin{equation} \label{eq:newfit}
M_{V} = (-2.94 \pm 0.06)\log P - (1.34 \pm 0.06).
\end{equation}
This is very similar to the line fitted by \citet{McNamara2011}.

\section{Discussion}

The results indicate that the dominant period in most \dsct\ stars lies on the ridge that coincides with the fundamental mode.  This does not necessarily mean that the dominant mode is the radial fundamental in every one of these stars, since it could also be a low-order nonradial mode.  Nevertheless, the diagram should prove useful when trying to identify the other modes in multi-periodic pulsators.  For example, three of the additional stars mentioned in Sec.~\ref{sec:additional} (cyan triangles in Figs.~\ref{fig:PL} and~\ref{fig:PL_2}) appear to fall into this class (Altair, 95~Vir and $\eta$~Ind).  In other stars (including the fourth additional star, $\beta$~Pic) the dominant mode has a period much shorter than the fundamental, corresponding to an overtone mode ($n>1$). 

In Fig.~\ref{fig:PL_2} we indicate three \kepler\ \dsct\ stars (black points with error bars) for which the dominant pulsation mode has been identified as the fundamental radial mode: 
KIC~5950759 (\citealt{Bowman2016}; $\log P = -1.15$),
KIC~2304168 (\citealt{Balona+Dziembowski2011}; $\log P = -0.91$), and
V2367~Cyg (KIC 9408694; \citealt{Balona++2012}; $\log P = -0.75$).
These stars all lie close to the main ridge, as expected.  Note that we have included error bars (see Sec.~\ref{sec:kepler-sample} for details) as an indication of the typical precision in $M_V$ for the \kepler\ sample.

In Fig.~\ref{fig:PL_2} we also show the two strongest modes in KIC~9700322, which were identified by \citet{Breger2011} as radial modes (black triangles).  The shorter-period mode, which has a slightly higher amplitude and is therefore the dominant mode, corresponds to the first overtone.  The longer-period mode (which should not really be plotted in the figure because it is not the dominant mode) is the radial fundamental and lies on the main ridge, as expected, while the period of the first overtone is a shorter by a factor of $0.78$ (shifted by $-0.11$ in $\log P$).

\subsection{Stars in the lower-right of the P--L diagram}
Some points in Fig.~\ref{fig:PL_2} lie well to the lower-right of the main ridge.  As discussed above, those from the \kepler\ sample are explained as harmonics in the Fourier spectrum of gravity modes (\gdor\ pulsations).  For the \citet{Rodriguez2000} catalogue (red squares), we have checked the seven stars that fall furthest to the lower-right in the diagram.  For four of these stars, subsequent publications have shown that the period in the catalogue is not from \dsct\ pulsations:
\begin{itemize}
    \item AD~Ari ($\log P = -0.57$, $M_V = 2.26$) is an ellipsoidal variable \citep{Handler+Shobbrook2002};
    \item $\epsilon$~UMa ($\log P = -0.90$, $M_V = 2.03$) is a rotating magnetic Ap star \citep{Shulyak2010};
    \item V1241~Tau ($\log P = -0.78$, $M_V = 2.06$) is an eclipsing binary \citep{Arentoft2004}; and
    \item V831~Her ($\log P = -0.94$, $M_V = 1.95$) is a constant star within the instability strip \citep{Henry2011}.
    \end{itemize}
These examples demonstrate the usefulness of the P--L diagram for identifying misclassified stars.  
For the other three stars, the situation is less clear:
\begin{itemize}
    \item BX~Scl (CS 22966$-$0043; $\log P = -1.43$, $M_V = 3.84$) is at the faint limit of our sample ($V=13.56$).  It is an SX~Phe pulsator and possibly an usual type of blue straggler \citep{Preston+Landolt1998}, which may explain its anomalous position.

    \item FP Ser (=40 Ser; $\log P = -0.70$, $M_V = 1.71$) has a period of 0.20\,d in the catalogue, which comes from \citet{Jackish1972}.  That paper actually suggests a period of 5--6 hours, based on observations over four nights in 1965, and it is clearly desirable to confirm this period.

    \item GS UMa ($\log P = -0.79$, $M_V = 1.66$) has a period of 0.164\,d from Hipparcos data, which was recently confirmed by \citet{Kahraman-Alicavus2018}.  However, those authors noted that GS~UMa is cooler than the red edge of the instability strip, and its position in Fig.~\ref{fig:PL_2} implies it may be a \gdor\ pulsator rather than a \dsct\ star.  

\end{itemize}
Data from the TESS Mission should help test these conclusions and clarify the classifications of these stars. 

\subsection{Interpretation of the P--L diagram}

\begin{figure*}
\includegraphics[width=0.75\linewidth]{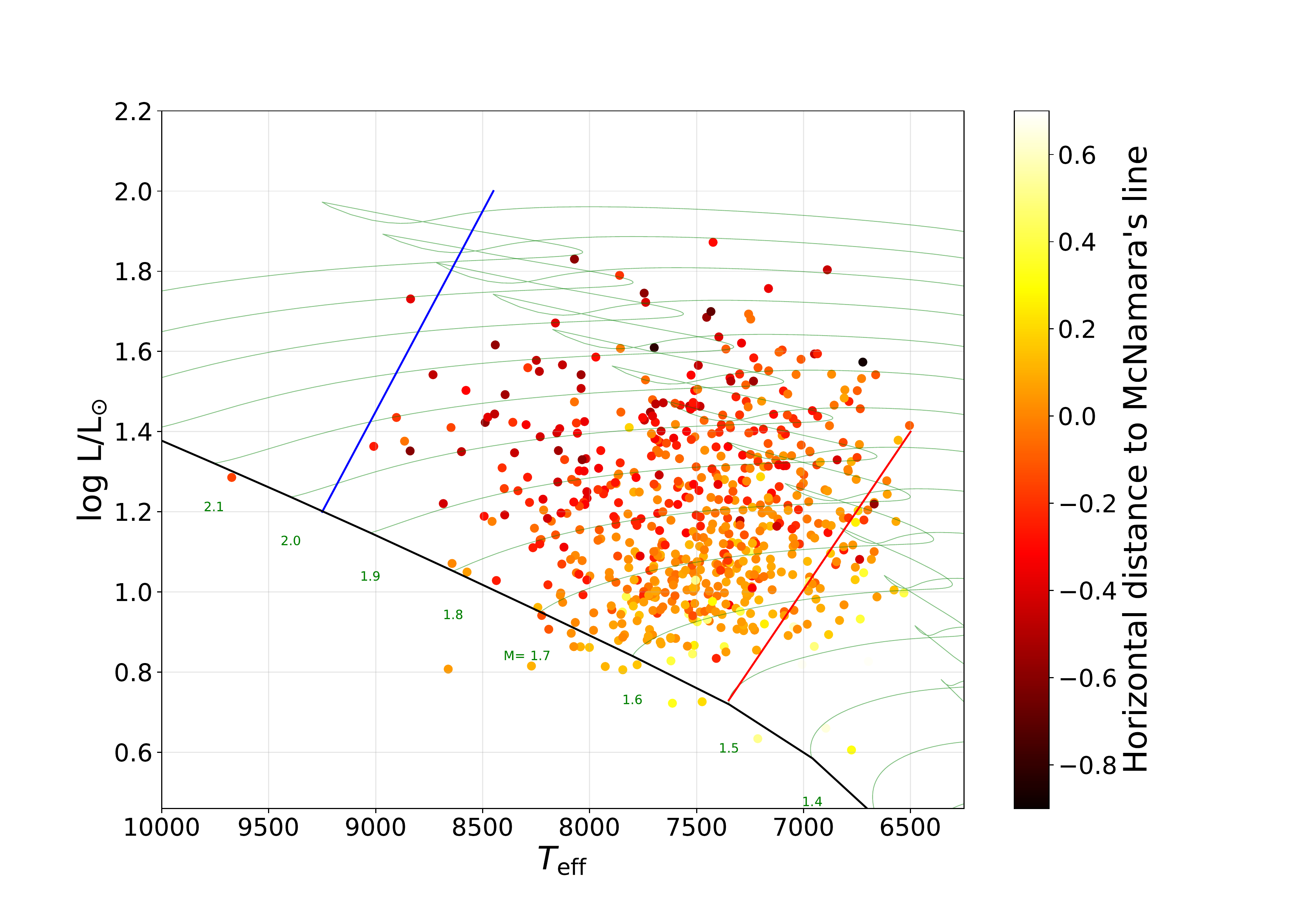}
\caption{HR diagram for our sample of \kepler\ \dsct\ stars having fractional parallax uncertainties less than 5 percent and pulsation semi-amplitudes above 1\,mmag. Luminosities and temperatures are taken from \citet{Murphy2019}. The colour indicates the horizontal distance from the fundamental-mode relation in the P--L diagram (Fig.~\ref{fig:histogram-highamp-offset}), with redder colours corresponding to shorter periods (higher pulsation overtones). The evolutionary tracks (green) are from \citet{Murphy2019} and have X = 0.71 and Z = 0.014. The blue and red lines are the theoretical instability strip boundaries using TDC and $\alpha_{\rm MLT} = 1.8$ from \citet{Dupret2005}.} 
\label{fig:hr}
\end{figure*}

Theoretical models by \citet{Dziembowski1997}, \citet{Houdek1999} and \citet{Dupret2005} indicate that the fundamental mode is unstable (i.e.\ excited) in stars on the cooler side of the instability strip, but that excitation shifts to progressively higher overtones at higher effective temperatures (see also \citealt{Xiong2001,Xiong2016}). To investigate this further, we show an HR diagram in Fig.~\ref{fig:hr} for the \kepler\ \dsct\ stars, where the colour indicates the horizontal distance from the fundamental P--L relation.  We clearly see that an increase in effective temperature correlates with the dominant period being shorter than the fundamental, which is consistent with a higher-order overtone being dominant.
This is also consistent with findings that \kepler\ \dsct\ stars with higher \Teff\ tend to have shorter dominant periods, i.e.\ a higher $\nu_{\rm max}$ \citep{Balona+Dziembowski2011, Barcelo-forteza2018}.

The puzzling finding from this work is the second ridge in the P--L diagram (Figs.~\ref{fig:PL} to~\ref{fig:histogram-highamp-offset}).  One possibility is that the second ridge is displaced vertically, in absolute magnitude, due to binarity. However, the vertical displacement is 0.9\,mag, higher than the 0.75\,mag expected for a binary sequence of equal-luminosity components. This difference is larger still, since \dsct\ stars preferentially have low-mass (low-luminosity) companions \citep{Murphy2018}. Also, the 72 stars in the \citet{Rodriguez2000} sample that are listed as binary systems by \citet{Liakos+Niarchos2017} do not preferentially lie on the second sequence.  
We consider it more likely that the second ridge is displaced horizontally, to shorter periods. The tendency of the shorter-period stars to have higher effective temperatures (Fig.~\ref{fig:hr})  is consistent with this, as discussed above.  However, the excess of stars having a dominant period that is half the fundamental needs an explanation.

Could the second ridge correspond to the second harmonic of the fundamental mode?  
It is common for \dsct\ stars to show combination frequencies, and a strong peak at frequency $f$ in the Fourier transform is often accompanied by a peak at twice that frequency.  However, in most cases the peak at $2f$ is much smaller in amplitude, and it would not appear in our P--L diagrams.

We are left to propose that the dominant period of the stars on the second ridge is an overtone that is excited to higher-than-expected amplitude due to a 2:1 resonance with the fundamental, even though the fundamental mode itself is not unstable.  Resonances are well known in pulsating stars \citep[e.g.,][]{Buchler1997,Kollath2011,Breger+Montgomery2014,Barcelo-forteza2015,Bowman2016}, and so it seems at least plausible that such a mechanism may boost the third or fourth overtone so that it becomes the dominant mode in cases where its frequency is twice that of the fundamental.  We encourage theoretical investigations of this possibility.

\section{Conclusions}

By using the {\em Gaia} DR2 parallaxes, we have examined the P--L relation for \dsct\ stars. 
In the current work, two samples of \dsct\ stars were constructed (see Sec.~\ref{sec:samples}), the first containing 228 stars from ground-based observations catalogued by \citet{Rodriguez2000} and the second including 1124 stars observed by \kepler. The absolute magnitude of each star was calculated by using {\em Gaia} DR2 parallaxes, including a correction for extinction, and then was plotted against the dominant period of the star (see Sec.~\ref{sec:method}).

Figure~\ref{fig:PL_2} shows the P--L relation for both samples, where only those \kepler\ stars with semi-amplitudes above 1\,mmag are included. In this figure, many stars fall in a ridge very close to the green dashed line, which corresponds to the published P--L relation of the radial fundamental mode in \dsct\ stars. 
The general distribution is in agreement with theoretical models, which indicate the excited mode in a hotter \dsct\ star would shift to higher overtones (see Fig.~\ref{fig:hr}).

There is an excess of stars in a second ridge for both samples, to the left of the main ridge by a distance of 0.3 in $\log P$, that could also be distinguished in histograms (Figs.~\ref{fig:histogram} and~\ref{fig:histogram-highamp-offset}). This ridge corresponds to stars having a dominant period that is half that of the main ridge. 
We suggest that this may be an excited overtone that is boosted by a 2:1 resonance with the fundamental. In future work, detailed examination of the Fourier Transforms of \kepler\ light curves could give us more information about the pulsation modes of the stars in the second ridge.

\section*{Acknowledgements}

We thank Radek Smolec and Pawel Moskalik for helpful discussions, and the referee for very useful comments.
We gratefully acknowledge support from the Australian Research Council, and
from the Danish National Research Foundation (Grant DNRF106) through its
funding for the Stellar Astrophysics Center (SAC).
This work has made use of data from the European Space Agency (ESA) mission {\em Gaia}, (\url{https://www.cosmos.esa.int/gaia}), 
processed by the {\em Gaia} Data Processing and Analysis Consortium (DPAC, \url{https://www.cosmos.esa.int/web/gaia/dpac/consortium}). Funding for the DPAC has been provided by national institutions, in particular the institutions participating in the {\em Gaia} Multilateral Agreement.
We are grateful to the entire Gaia and \kepler\ teams for providing the data used in this paper.


\bibliographystyle{mnras}
\bibliography{references}


\bsp	
\label{lastpage}
\end{document}